\documentclass[3p,preprint,11pt]{elsarticle}

\journal{Communications in Nonlinear Science and Numerical
Simulation}

\usepackage{amssymb,amsmath,amsthm,bm}

\newtheorem{lemma}{Lemma}
\newtheorem{proposition}{Proposition}

\biboptions{sort&compress}

\usepackage{graphicx}
\graphicspath{{figures/}}

\newlength\imagewidth
\usepackage{color}
\definecolor{dgreen}{rgb}{0,.6,0}
\usepackage[normalem]{ulem}

\begin{document}

\begin{frontmatter}

\title{Breaking a new substitution-diffusion based image cipher using chaotic standard and logistic maps}

\author[hk-polyu]{Chengqing Li\corref{corr}}
\ead{zjulcq@gmail.com}
\author[gm-constan]{Shujun Li\corref{corr}}
\ead[URL]{www.hooklee.com}
\author[hk-polyu]{Kwok-Tung Lo}
\cortext[corr]{Corresponding authors.}

\address[hk-polyu]{Department of Electronic and Information Engineering, The Hong Kong
Polytechnic University, Hong Kong, China}

\address[gm-constan]{Fachbereich Informatik und Informationswissenschaft,
Universit\"at Konstanz, Fach M697, Universit\"atsstra{\ss}e 10,
78457 Konstanz, Germany}

\begin{abstract}

Recently, an image encryption scheme based on chaotic standard and
logistic maps was proposed. This paper studies the security of the
scheme and shows that it can be broken with only one
chosen-plaintext. Some other security defects of the scheme are also
reported.
\end{abstract}


\end{frontmatter}

\section{Introduction}

With the rapid development of information technology, multimedia
data are transmitted over all kinds of wired/wireless networks more
and more frequently. Consequently, the security of multimedia data
becomes a serious concern of many people. However, the traditional
text encryption schemes can not be used in a naive way to protect
multimedia data efficiently in some applications, mainly due to the
big differences between textual and multimedia data and some special
requirements of the whole multimedia system. This challenge stirs
the design of special multimedia encryption schemes to become a hot
research topic in multimedia signal processing area in the past
decade. Because of the subtle similarity between chaos and
cryptography, a great number of multimedia encryption schemes based
on chaos have been presented \cite{Chen&Yen:RCES:JSA2003,
YaobinMao:CSF2004, Flores:EncryptLatticeChaos06,
Tong:ImageCipher:IVC07}. Unfortunately, many of them have been found
to have security problems from the cryptographical point of view
\cite{Li:AttackingMaoScheme2007, Li:AttackingRCES2008,
David:AttackingChaos08, Goce:cryptanalysis:TM08,
Li:BreakImageCipher:IVC09}. Some general rules about evaluating
security of chaos-based encryption schemes can be found in
\cite{AlvarezLi:Rules:IJBC2006,
Li:ChaosImageVideoEncryption:Handbook2004}.

Since 2003, Pareek et al. have proposed a number of different
encryption schemes based on one or more chaotic maps
\cite{Pareek:PLA2003, Pareek:CNSNS2005, Pareek:ImageEncrypt:IVC2006,
Pareek:CNSNS2009}. Recent cryptanalytic results
\cite{Alvarez:PLA2003, Li:AttackingCNSNS2008, Li:AttackingIVC2009}
have shown that all the three schemes proposed in
\cite{Pareek:PLA2003, Pareek:CNSNS2005, Pareek:ImageEncrypt:IVC2006}
have security defects. In \cite{Pareek:CNSNS2009}, a new image
encryption scheme based on the Logistic and standard maps was
proposed, where the two maps are used to generate a pseudo-random
number sequence (PRNS) controlling two kinds of encryption
operations. The present paper focuses on a re-evaluation of the
security of this new scheme, and reports the following findings: 1)
the scheme can be broken with only one chosen image; 2) there are
also some other security defects of the scheme.

The rest of this paper is organized as follows.
Section~\ref{sec:encryptscheme} introduces the image encryption
scheme under study briefly. Our cryptanalytic results are presented
in Sec.~\ref{sec:cryptanalysis} in detail. The last section
concludes the paper.

\section{The image encryption scheme under study}
\label{sec:encryptscheme}

The plaintext encrypted by the image encryption scheme under study
is a RGB true-color image of size $M\times N$ (height$\times$width),
which can be denoted by an $M\times N$ matrix of 3-tuple pixel
values $\bm{I}=\{(R(i,j), G(i,j), B(i,j))\}_{0\leq i\leq M-1 \atop
0\leq j\leq N-1}$. Denoting the cipher image by $\bm{I}'=\{(R'(i,j),
G'(i,j), B'(i,j))\}_{0\leq i\leq M-1 \atop 0\leq j\leq N-1}$, the
image encryption scheme can be described as follows\footnote{To make
the presentation more concise and complete, some notations in the
original paper are modified.}.

\begin{itemize}
\item
\textit{Secret key}: three floating-point numbers $x_0$, $y_0$, $K$,
and one integer $L$, where $x_0$, $y_0\in(0, 2\pi)$, $K>18$,
$100<L<1100$.

\item
\textit{Initialization}: prepare data for encryption/decryption by
performing the following steps.
\begin{itemize}
\item[a)]
Generate four XORing keys as follows: $\textit{Xkey}(0)=\lfloor
256x_0/(2\pi)\rfloor$, $\textit{Xkey}(1)=\lfloor
256y_0/(2\pi)\rfloor$, $\textit{Xkey}(2)=\lfloor K\bmod 256
\rfloor$, $\textit{Xkey}(3)=(L \bmod 256)$.

\item[b)]
Iterate the standard map Eq.~(\ref{eq:standardmap}) from the initial
conditions $(x_0,y_0)$ for $L$ times to obtain a new chaotic state
$(x_0', y_0')$. Then, further iterate it for $MN$ more times to get
$MN$ chaotic states $\{(x_i,y_i)\}_{i=1}^{MN}$.

\begin{equation}
\label{eq:standardmap}
\begin{cases}
x  =  (x+K\sin(y))\bmod(2\pi),\\
y  =  (y+x+K\sin(y))\bmod(2\pi),
\end{cases}
\end{equation}

\item[c)]
Iterate the Logistic map Eq.~(\ref{eq:logistic}) from the initial
condition $z_0=((x_0'+y_0')\bmod 1)$ for $L$ times to get a new
initial condition $z_0'$. Then, further iterate it for $MN$ times to
get $MN$ chaotic states $\{z_i\}_{i=1}^{MN}$.

\begin{equation}
\label{eq:logistic} z=4z(1-z).
\end{equation}

\item[d)]
Generate a chaotic key stream (CKS) image
$\bm{I}_{CKS}=\{(\textit{CKSR}(i,j), \textit{CKSG}(i,j),$
$\textit{CKSB}(i,j))\}_{0\leq i\leq M-1 \atop 0\leq j\leq N-1}$ as
follows: $\textit{CKSR}(i,j)=\left\lfloor
256x_k/(2\pi)\right\rfloor$, $\textit{CKSG}(i,j)=\left\lfloor
256y_k/(2\pi)\right\rfloor$ and $\textit{CKSB}(i,j)=\left\lfloor
256z_k\right\rfloor$, where $k=iN+j+1$.
\end{itemize}

\item
\textit{Encryption procedure}: a simple concatenation of the
following four encryption operations.

\begin{itemize}
\item
\textit{Confusion I}: masking the plain pixel values by the four
XORing keys $\{\textit{Xkey}(i)\}_{i=0}^3$.

For $k=0,\ldots,MN-1$, do the following masking operations.
\begin{eqnarray}
R^{\star}(i,j) & = & R(i,j)\oplus \textit{Xkey}( (3\cdot k )\bmod 4 ),\\
G^{\star}(i,j) & = & G(i,j)\oplus \textit{Xkey}( (3\cdot k+1)\bmod 4),\\
B^{\star}(i,j) & = & B(i,j)\oplus \textit{Xkey}( (3\cdot k +2)\bmod
4 ),
\end{eqnarray}
where $i=\lfloor k/N\rfloor$, $j=(k\bmod N)$.

\item
\textit{Diffusion I}: scanning all pixel values from the first one
row by row (from top to bottom), and masking each pixel (except for
the first scanned pixel) by its predecessor in the scan.

Set $R^{*}(0, 0)=R^{\star}(0, 0)$, $G^{*}(0, 0)=G^{\star}(0, 0)$,
$B^{*}(0, 0)=B^{\star}(0, 0)$. For $k=1,\ldots,MN-1$,
\begin{eqnarray}
R^{*}(i, j) & = &  R^{\star}(i, j) \oplus R^{*}(i', j'),\\
G^{*}(i, j) & = &  G^{\star}(i, j) \oplus G^{*}(i', j'),\\
B^{*}(i, j) & = &  B^{\star}(i, j) \oplus B^{*}(i', j'),
\end{eqnarray}
where $i=\lfloor k/N\rfloor$, $j=(k\bmod N)$, $i'=\lfloor
(k-1)/N\rfloor$ and $j'=((k-1)\bmod N)$.

\item
\textit{Diffusion II}: scanning all pixel values from the last one
column by column (from right to left), and masking each pixel
(except for the first scanned pixel) by its predecessor in the scan.

Set $R^{**}(M-1, N-1)=R^{*}(M-1, N-1)$, $G^{**}(M-1, N-1)=G^{*}(M-1,
N-1)$, $B^{**}(M-1, N-1)=B^{*}(M-1, N-1)$. For $k=MN-2,\ldots,0$,
\begin{eqnarray}
R^{**}(i, j) & = & R^{*}(i, j)\oplus G^{**}(i',j')\oplus B^{**}(i',j'),\\
G^{**}(i, j) & = & G^{*}(i, j)\oplus B^{**}(i',j')\oplus R^{**}(i',j'),\\
B^{**}(i, j) & = & B^{*}(i, j)\oplus R^{**}(i',j')\oplus
G^{**}(i', j'),
\end{eqnarray}
where $i=(k\mod M)$ and $j=\lfloor k/M \rfloor$, $i'=((k+1)\mod M)$,
$j'=\lfloor (k+1)/M \rfloor$.

\item
\textit{Confusion II}: masking the pixel values with the CKS image
pixel by pixel.

For $k=0,\ldots,MN-1$,
\begin{eqnarray}
R'(i,j) & = & R^{**}(i,j)\oplus \textit{CKSR}(i,j),\\
G'(i,j) & = & G^{**}(i,j)\oplus \textit{CKSG}(i,j),\\
B'(i,j) & = & B^{**}(i,j)\oplus \textit{CKSB}(i,j).
\end{eqnarray}
where $i=\lfloor k/N\rfloor$, $j=(k\bmod N)$.
\end{itemize}

\item
\textit{Decryption procedure}: the simple reversion of the above
encryption procedure.
\end{itemize}

\section{Cryptanalysis}
\label{sec:cryptanalysis}

\subsection{A chosen-plaintext attack}

In the chosen-plaintext attack, the attacker can choose plaintexts
arbitrarily and obtain the corresponding ciphertexts. The goal of
the attack is to gain some further information which helps reveal
the other plaintexts encrypted with the same secret key. For the
image encryption scheme under study, an equivalent version of the
secret key can be reconstructed easily from only one pair of
chosen-plaintext as shown in Proposition~1.

\begin{lemma}
Let $\bar{\bm{I}}^{**}$ denote the encryption result of $\bm{I}$
without performing the two confusion steps. Then,
$\bm{I}'=\bar{\bm{I}}^{**}\oplus\bar{\bm{I}}_{\textit{Xkey}}^{**}\oplus\bm{I}_{\textit{CKS}}$.
\end{lemma}
\begin{proof}
After the first confusion step,
$\bm{I}^{\star}=\bm{I}\oplus\bm{I}_{\textit{Xkey}}$, where
$\bm{I}_{\textit{Xkey}}$ is the pseudo-image composed of the four
XORing keys. Observing the operations involved in the two diffusion
steps, we can see both steps can be performed on $\bm{I}$ and
$\bm{I}_{\textit{Xkey}}$ separately and XOR the results, which means
that
$\bm{I}^{**}=\bar{\bm{I}}^{**}\oplus\bar{\bm{I}}_{\textit{Xkey}}^{**}$.
Then, after performing the last confusion step, we have
$\bm{I}'=\bar{\bm{I}}^{**}\oplus\bar{\bm{I}}_{\textit{Xkey}}^{**}\oplus\bm{I}_{\textit{CKS}}$,
which proves this lemma.
\end{proof}

\begin{proposition}
If $\bm{I}_0$ is a zero image, i.e., $\bm{I}_0=\bm{0}$, then
$\bm{I}'=\bar{\bm{I}}^{**}\oplus\bm{I}_0'$.
\end{proposition}
\begin{proof}
This is a straightforward result of the fact
$\bm{I}_0^{\star}=\bm{0}\oplus\bm{I}_{\textit{Xkey}}=\bm{I}_{\textit{Xkey}}$.
\end{proof}

In case $\bm{I}_0'$ is known, the above proposition means that the
plain-image $\bm{I}$ can be recovered from $\bm{I}'$ by the
following steps: 1) $\bar{\bm{I}}^{**}=\bm{I}'\oplus\bm{I}_0'$; 2)
perform the two diffusion steps on $\bar{\bm{I}}^{**}$ in an inverse
order, which exactly recovers $\bm{I}$. In other words, by taking
$\bm{I}_0$ as a chosen-image, we can get an equivalent key
$\bm{I}_0'$ to decrypt any cipher-image encrypted with the same
secret key $(x_0,y_0,K,L)$.

\begin{figure}[!htb]
\centering
\begin{minipage}[t]{\imagewidth}
\centering
\includegraphics[width=\imagewidth]{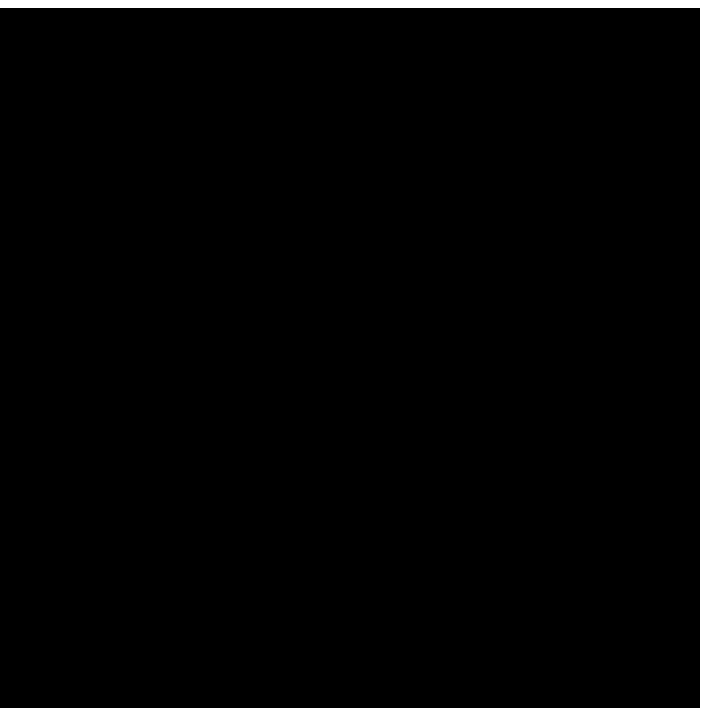}
a)
\end{minipage}
\begin{minipage}[t]{\imagewidth}
\centering
\includegraphics[width=\imagewidth]{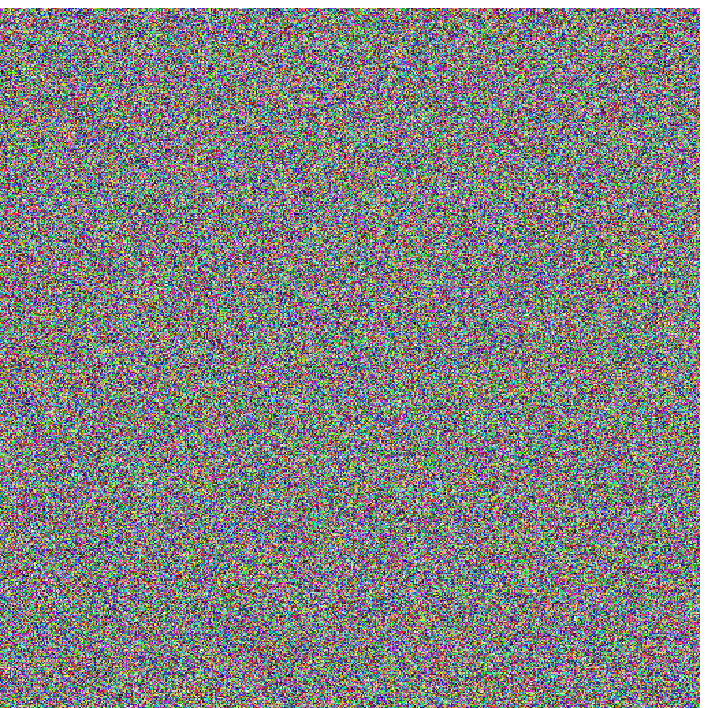}
b)
\end{minipage}
\begin{minipage}[t]{\imagewidth}
\centering
\includegraphics[width=\imagewidth]{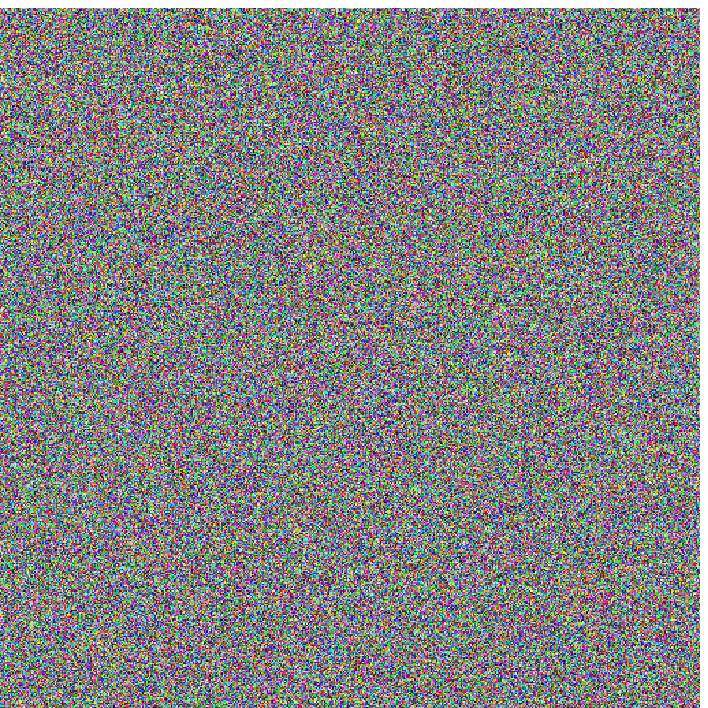}
c)
\end{minipage}
\begin{minipage}[t]{\imagewidth}
\centering
\includegraphics[width=\imagewidth,height=\imagewidth]{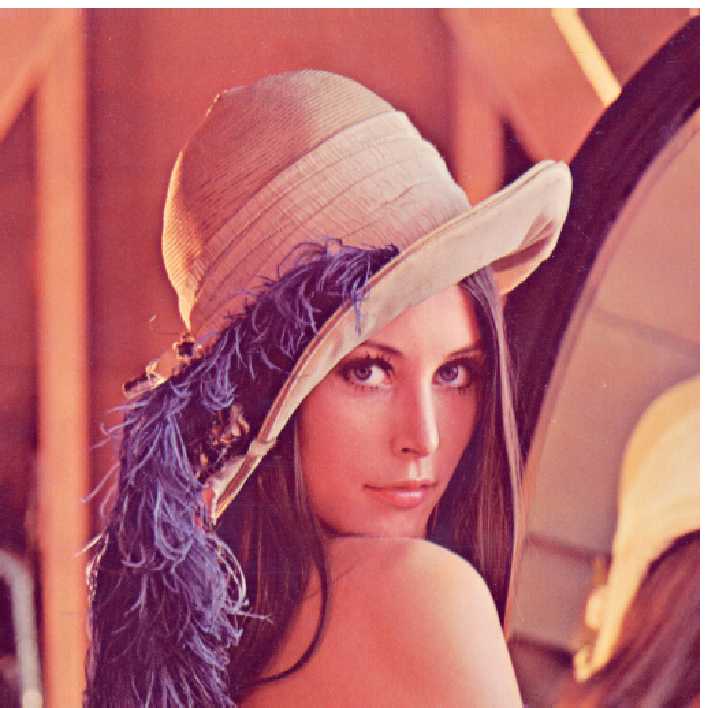}
d)
\end{minipage}
\caption{An experimental result of the proposed chosen-plaintext
attack: a) the chosen plain-image $\bm{I}_0=\bm{0}$; b) the
equivalent key $\bm{I}_0'$; c) a cipher-image encrypted with the
same key; d) the recovered plain-image ``Lenna''.}
\label{figure:chosenplaintextattack}
\end{figure}

We have performed some experiments to verify the correctness of the
above chosen-plaintext attack. With the secret key $(x_0, y_0, K,
L)=(3.98235562892545,$ $ 1.34536356538912, 108.54365761256745,$ $
110)$, the equivalent key $\bm{I}_0'$ was constructed from the zero
image, which are shown in Figs.~\ref{figure:chosenplaintextattack}a
and b, respectively. Then, $\bm{I}_0'$ was used to recover the
cipher-image shown in Fig.~\ref{figure:chosenplaintextattack}c, and
successfully recovered the plain-image ``Lenna''
(Fig.~\ref{figure:chosenplaintextattack}d). \color{black}

\subsection{Some other security problems}

\subsubsection{Insufficient randomness of the PRNS $\{\textit{CKSB}(k)\}$}

As illustrated in \cite{Li:AttackingBitshiftXOR2007}, the randomness
of the pseudo-random bit sequence derived from chaotic states
generated by iterating Logistic map is very weak. To further verify
the randomness of the PRNS $\{\textit{CKSB}(k)\}$ generated via the
Logistic map of fixed control parameter, the NIST statistical test
suite \cite{Rukhin:TestPRNG:NIST} was employed to test the
randomness of 100 PRNSes of length $512\cdot 512=262144$ (the number
of bytes used for encryption of a $512\times 512$ plain color
image). Note that the 100 sequences were generated with randomly
selected secret keys. For each test, the default significance level
0.01 was used. The results are shown in Table~\ref{table:test}, from
which one can see that the PRNS $\{\textit{CKSB}(k)\}$ is not random
enough.

\begin{table}[!htbp]
\centering\caption{The performed tests with respect to a
significance level 0.01 and the number of sequences passing each
test in 100 randomly generated sequences.}\label{table:test}
\begin{tabular}{c|c}
\hline Name of Test                                    & Number of Passed Sequences\\
\hline\hline Frequency                                 & 95 \\
\hline Block Frequency ($m=100$)                       & 0  \\
\hline Cumulative Sums-Forward                         & 93 \\
\hline Runs                                            & 0  \\
\hline Rank                                            & 0  \\
\hline Non-overlapping Template ($m=9$, $B=010000111$) & 10 \\
\hline Serial ($m=16$)                                 & 0  \\
\hline Approximate Entropy ($m=10$)                    & 0  \\
\hline FFT                                             & 0  \\
\hline
\end{tabular}
\end{table}

\subsubsection{Insensitivity with respect to change of plaintext}

In \cite[Sec. 5.5]{Pareek:CNSNS2009}, it is recognized that the
sensitivity of cipher-image generated by an image encryption scheme
with respect to change of plain-image is very important, but the
image encryption scheme under study is actually very far from the
desired property. As well known in cryptography, this property is
termed as avalanche effect. Ideally, it requires the change of any
single bit of plain-image will make every bit of cipher-image change
with a probability of one half. However, the image encryption scheme
under study can not satisfy this property due to the following
points.

\begin{itemize}
\item
Only one kind of operation (XOR) is involved in the whole scheme;

\item
Any bit of plain-image only influences the bits at the same level in
the cipher-image;

\item
Any pixel of plain-image does not influence other pixels in the
cipher-image uniformly.
\end{itemize}

\begin{figure}[!htb]
\centering
\begin{minipage}[t]{0.7\imagewidth}
\centering
\includegraphics[width=0.7\imagewidth]{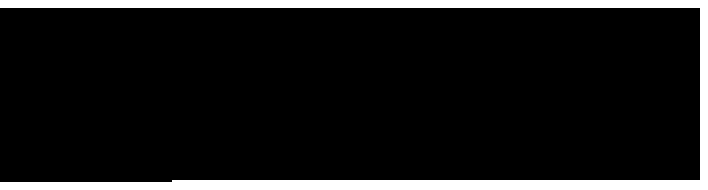}
a)
\end{minipage}
\begin{minipage}[t]{0.7\imagewidth}
\centering
\includegraphics[width=0.7\imagewidth]{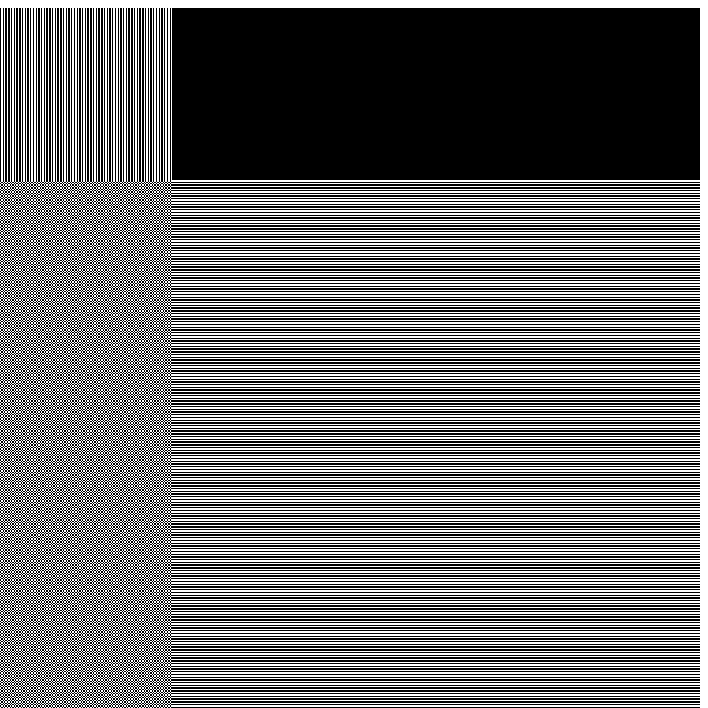}
b)
\end{minipage}
\begin{minipage}[t]{0.7\imagewidth}
\centering
\includegraphics[width=0.7\imagewidth]{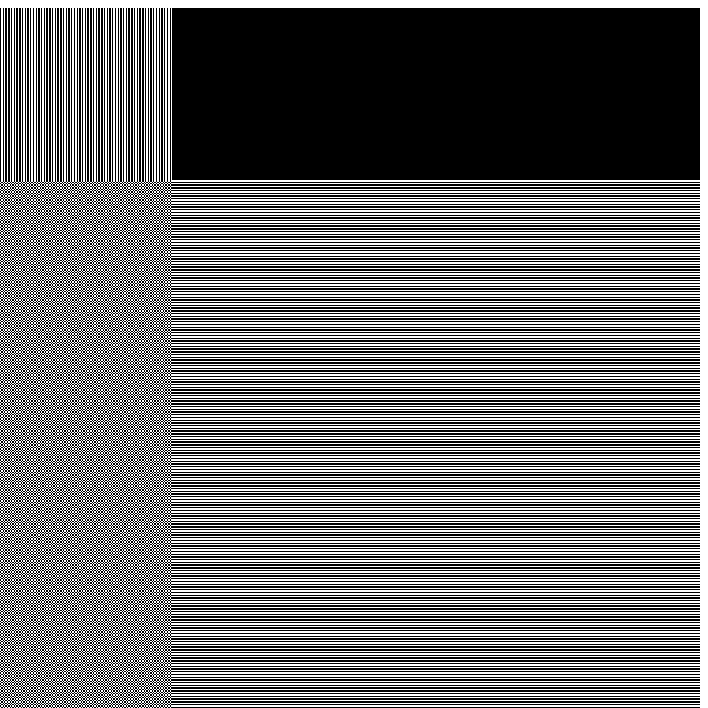}
c)
\end{minipage}
\caption{The locations of changed bits in different color channels
of the cipher-image, when the 6-th bit of the red component of the
pixel at location (127, 127) in the plain-image was changed: a) red
channel; b) green channel; c) blue
channel.}\label{figure:changebits}
\end{figure}

To show this defect clearly, we made an experiment by changing only
one bit of the red channel of the plain-image shown in
Fig.~\ref{figure:chosenplaintextattack}d. It is found that only some
bits at the same level in the corresponding cipher-image were
changed. The locations of the changed bits are shown in
Fig.~\ref{figure:changebits}, in which the white dots denote changed
locations and black ones denote unchanged ones.

\section{Conclusion}

In this paper, the security of a new image encryption scheme based
on two chaotic maps is analyzed in detail. It is found that the
scheme can be broken with only one chosen plain-image. In addition,
some other security defects about randomness of a PRNS involved, and
sensitivity with respect to change of plain-image are also reported.
Due to such a low level of security, we recommend not to use the
image encryption scheme under study in any serious applications.

\section*{Acknowledgement}

Chengqing Li was supported by The Hong Kong Polytechnic University's
Postdoctoral Fellowships Scheme under grant no. G-YX2L. Shujun Li
was supported by a fellowship from the Zukunftskolleg of the
Universit\"at Konstanz, Germany, which is part of the
``Exzellenzinitiative'' Program of the DFG (German Research
Foundation). The work of Kowk-Tung Lo was supported by the Research
Grant Council of the Hong Kong SAR Government under Project 523206
(PolyU 5232/06E).

\bibliographystyle{elsarticle-num}
\bibliography{Pareek}

\end{document}